\documentclass[prb,reprint,showpacs]{revtex4-1}
\usepackage{amsmath}
\usepackage{graphicx}
\usepackage{amssymb}

\begin{document}
\title{Electronic thermal conductivity as derived by density functional theory}
\author {M. X.\ Chen}
\affiliation{Department of Physics, University of Wisconsin-Milwaukee, Milwaukee, Wisconsin 53211, USA}
\author {R.\ Podloucky}
\affiliation{Department of Physical Chemistry, University of Vienna,
Sensengasse
8/7, 1090 Vienna, Austria}

\date{\today}
\pacs{1,2,3}

\begin{abstract}
Reliable evaluation of the lattice thermal conductivity is of importance for
optimizing the figure-of-merit of thermoelectric materials.  Traditionally,
when deriving the phonon mediated thermal conductivity $\kappa_{ph} = \kappa -
\kappa_{el}$ from the measured total thermal conductivity $\kappa$ the
constant Lorenz number $L_0$ of the Wiedemann-Franz law
\mbox{$\mathbf{\kappa_{el}}=T L_0 \sigma$} is chosen.  The present
study demonstrates that this procedure is not reliable when the Seebeck
coefficient $|S|$ becomes large which is exactly the case for a thermoelectric
material of interest.  Another approximation using $L_0-S^2$,
which seem to work better for medium values of $S^2$ also fails when $S^2$ becomes large, 
as is the case when the system becomes semiconducting/insulating. 
For a reliable estimation of $\kappa_{el}$ it is
proposed, that a full first-principles calculations by combining 
density functional theory with
Boltzmann's transport theory has to be made. For the present
study such an approach was
chosen for investigating the clathrate type-I compound
Ba$_8$Au$_{6-x}$Ge$_{40+x}$ for a series of dopings or compositions $x$. 
For a doping of $0.8$ electrons corresponding 
to $x=0.27$ the calculated temperature
dependent Seebeck coefficient agrees well with recent experiments
corroborating the validity of the density functional theory approach.
\end{abstract}

%71.20.-b         Electron density of states and band structure of crystalline solids
%72.15.Eb         Electrical and thermal conduction in crystalline metals and alloys
%72.15.Jf         Thermoelectric and thermomagnetic effects

\maketitle
Thermal conductivity plays an important role  for the thermoelectric
performance of a material as expressed by the figure-of-merit 
$ZT = T$S$^2\sigma /(\kappa_{el}+\kappa_{ph})$ which includes the  Seebeck coefficient $S$, the
electrical conductivity $\sigma$, and the thermal conductivity
$\kappa=\kappa_{el} + \kappa_{ph}$  summing up the contributions of electronic
states and phonon mediated processes.
Consequently, a low  thermal conductivity 
in combination with large values of $S$ and $\sigma$ are desirable in order to
achieve large values of $ZT$.  Considerable efforts for lowering $\kappa$ 
by reducing $\kappa_{ph}$ were made by utilizing
structural properties, such as building up superlattices
\cite{venkatasubramanian_2001,beyer_pbte_2002,caylor_enhanced_2005,boettner_aspects_2006}
and incorporating suitable filler atoms into structural cages
\cite{sales_filled_1996,morelli_low_1995,nolas_effect_1996,sales_thermoelectric_2000,
nolas_high_2000,nolas_skutterudites:_1999,lamberton_high_2002,
nolas_semiconducting_1998}. These concepts
rely on the strong scattering of heat-transporting phonon modes.
However, neither $\kappa_{el}$ nor $\kappa_{ph}$ are directly measured.
Rather, $\kappa_{ph}$ is derived by subtracting $\kappa_{el}$ from the
measured total thermal conductivity, i.e., $\kappa_{ph} \approx \kappa^{meas.} - \kappa_{el}$
in which  the electronic thermal conductivity is estimated via the
Wiedemann-Franz (WF) relation for simple metals, $\kappa_{el} \approx T $L$_0 \sigma$
\cite{morelli_low_1995,nolas_effect_1996,sales_thermoelectric_2000,nolas_high_2000,nolas_skutterudites:_1999,
lamberton_high_2002,
ohtaki_hightemperature_1996,li_high_1999,takahata_low_2000,toprak_impact_2004,
androulakis_nanostructuring_2006,tang_preparation_2007,
zhou_nanostructured_2008,shi_low_2008,li_high_2009}.
In this expression, L$_0$ is a universal constant and does not depend on temperature and
materials properties.  In  the present work it is shown
by a density functional theory (DFT) study for a typical thermoelectric material 
that the application of the WF law leads to unreliable estimates of $\kappa_{el}$ in particular when the
Seebeck coefficient of the material is large, which is exactly the case of interest.

The present theoretical study is based on the same DFT concept
as applied for first-principles calculations of Seebeck coefficients
(for example, see Ref. \onlinecite{zeiringer_2011}). In the present work the WF
law is generalized by introducing a material and temperature dependent Lorenz tensor $\mathbf{L}$,
for which Boltzmann's transport theory  in combination with 
electronic properties derived by DFT calculation is used. This procedure is
applied for the clathrate type-I compound Ba$_8$Au$_{6-x}$Ge$_{40+x}$, which is
a prototypical thermoelectric material and for which also very recent measurements
of Seebeck coefficients are available enabling a  test of the validity of the
present theoretical approach.

For the present purpose, the WF law is generalized to
\begin{equation}
\mathbf{\kappa_{el}} = T \mathbf{L \sigma}~,
\label{kappa_wf}
\end{equation}
in which the tensor $\mathbf{\kappa_{el}}$ is linearly related to the conductivity tensor
$\mathbf{\sigma}$ (as defined in Eq.~\ref{electrical_conductivity}) via the Lorenz tensor $\mathbf{L}$. 
These quantities as well as the Seebeck tensor $\mathbf{S}$ (Eq.~\ref{seebeck}) are second rank tensors
\cite{nye_tensor}.  In accordance with Boltzmann's transport theory one
derives \cite{tritt_thermal_2004}
\begin{equation}
\mathbf{\kappa_{el}} =
\frac{1}{T}\left(\mathbf{K_2}-\mathbf{K_1}^2\mathbf{K_0}^{-1}\right) 
\label{kappa_boltzmann}
\end{equation}
for which  Eq.~\ref{kn} is utilized for the definition of the tensors
$\mathbf{K_n}$.
The Lorenz tensor $\mathbf{L}$ can now be formulated as
\begin{equation}
\mathbf{L} = \mathbf{L_1} - \mathbf{S}^2
\label{L_tensor} 
\end{equation}
whereby $\mathbf{L_1}$ is expressed as \cite{tritt_thermal_2004} 
\begin{equation}
\mathbf{L_1} = \frac{1}{e^2T^2}\mathbf{K_2 K_0}^{-1}~.
\label{L1_tensor}
\end{equation}

For a free-electron like metal the second term at the right hand side of
Eq.~\ref{L_tensor}  is negligible, because $|S|$ is small.  
It is then obvious that
the deviation from the free-electron like behavior is caused by the Seebeck
coefficient in terms of $-\mathbf{S}^2$. At low temperatures the original WF
law is a reasonable approximation, i.e. $\mathbf{L_1} \approx \mathbf{L_0}$,
whereby $\mathbf{L_0}$ would be a tensor with constant coefficients  
L$_0=\frac{\pi^2k_B^2}{3} = 2.44 \times 10^{-8} W\Omega/K^2$,  involving
Boltzmann's constant $k_B$.

When assuming a constant relaxation time $\tau=const$
--as it is the standard approach for first-principles calculations of the
Seebeck coefficients (for example, see Ref.~\onlinecite{zeiringer_2011})-- then
$\tau$ cancels out in the components of $\mathbf{L}$ 
since it appears in the numerator as well as in the denominator
in  Eqs.~\ref{L_tensor} and \ref{L1_tensor}.
The same holds for the Seebeck coefficients because of  Eq.~\ref{seebeck}. 
For the following discussion it should be noted that the crystal structure of the material under study
is of cubic symmetry. As a consequence of this high symmetry
all second rank tensors are diagonal
and the three diagonal coefficients are equal. Therefore, only one coefficient needs
to be considered for each tensor.  However, the derivations and
calculations can be done for a general crystal symmetry and for tensors with
less symmetry and more components. If the tensors are symmetry averaged (as
needed for a polycrystalline material), again only one
coefficient needs to be considered.

\begin{figure}
     \includegraphics[width=0.44\textwidth]{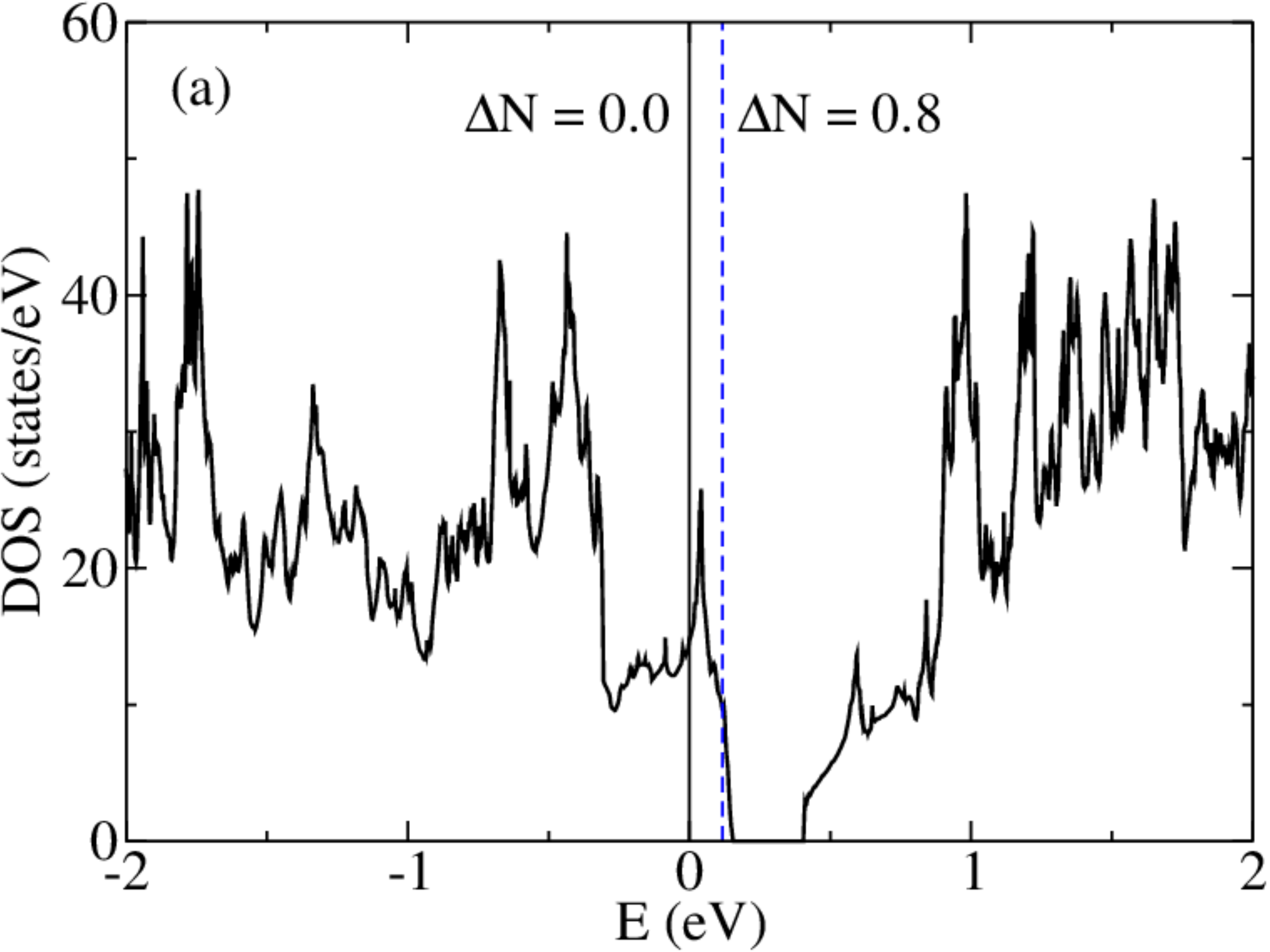} \vspace*{0.2cm} \\
     \includegraphics[width=0.44\textwidth]{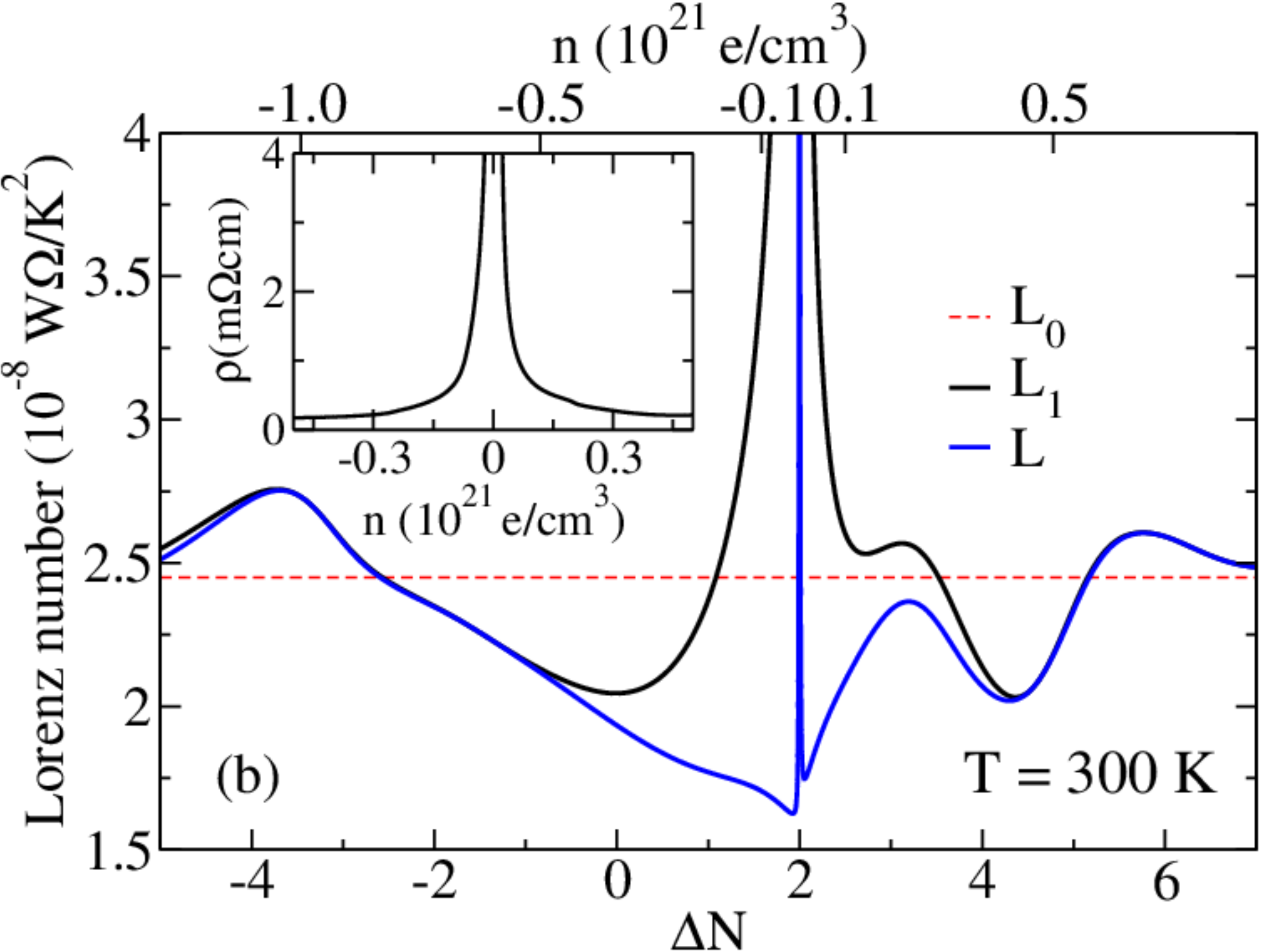}
\caption{
DFT results for Ba$_8$Au$_{6}$Ge$_{40}$:    
(a) density of states (DOS) vs. energy $E$. 
The Fermi levels of the undoped ($E=0$ for $x=0$) and the electron doped
($x=0.27, \Delta$N = 0.8) compound are
indicated by full and dashed lines, respectively. Doping was modelled within
the rigid band approximation.
The type-I clathrate structure is cubic with  space group \textit{Pm-3n}.
The relaxed cubic lattice parameter is $a=10.70$ \AA{},
and the site specific coordinates are
 2\textit{a} (0,0,0) and 6\textit{c} (0.25,0.5,0) for Ba, 6\textit{d}
(0.25,0.5,0) for Au, 16\textit{i} (0.183, 0.183, 0.183) and 24\textit{k} (0, 0.117, 0.309) for Ge.
(b) constant Lorenz number $L_0$ (dashed red line), coefficient $L$ of the
cubic Lorenz tensor (solid blue line) and
its approximation L$_1$ (solid black line) (Eqs. \ref{L_tensor}, \ref{L1_tensor}) at 
300 K vs. rigid-band doping $\Delta N$ (lower abscissa) or carrier
concentration (upper abscissa) $n$. 
Negative/positive values of $\Delta N$ or $n$ correspond to hole/electron
doping or carriers. 
For $\Delta N = 2$ the Fermi energy falls
into the gap of the DOS.
The inset shows the carrier-concentration dependent behavior of the resistivity
$\rho$, the inverse of the conductivity $\sigma$ as calculated according to 
Eq.~\ref{electrical_conductivity} for a constant relaxation time $\tau$ = 1 $\times$ 10$^{-14}$ s.
}
    \label{fig1}
  \end{figure}

What remains to be done
is the DFT calculation of the electronic structure of the actual material, 
Ba$_8$Au$_6$Ge$_{40}$.
For that purpose, the Vienna \textit{Ab initio} Simulation Package (VASP)
\cite{kresse_efficiency_1996,kresse_efficient_1996} was used for which the
pseudopotentials were constructed according to the projector augmented wave method 
\cite{bloechl_projector_1994,kresse_ultrasoft_1999}.
The exchange-–correlation functional was parametrized in terms of the
local density approximation according to Ceperley and Alder
\cite{ceperley_ground_1980}. 
The valence state configuration for the construction of the pseudopotentials included the 5s, 5p and
6s atomic states for Ba, the 6s and 5d states for Au, and the 3d,
4s and 4p states for Ge. For the Brillouin zone integration  a 5 $\times$ 5
$\times$ 5 grid of $\mathbf{k}$-points was found to be sufficiently accurate
concerning the relaxed structural parameters as mentioned
in the caption of Fig.~\ref{fig1}(a).  
Electronic transport properties were derived by utilizing the Boltzmann transport
equations within the constant relaxation time approximation as implemented in the program package BoltzTrap
\cite{zeiringer_2011,madsen_boltztrap._2006, chen_thesis_2012}. 
For that purpose  the Kohn-Sham eigenvalues $\varepsilon_i(\mathbf{k})$ were determined
on a very dense  25 $\times$ 25 $\times$ 25 grid of $\mathbf{k}$-points.
For the investigation of the thermoelectric properties of 
off-stoichiometric compounds Ba$_8$Au$_{6-x}$Ge$_{40+x}$
with varying $x$ or number of valence electrons (i.e. doping $\Delta N$)
the rigid-band construction was used which consists in
shifting the Fermi energy according to the doping  without changing the
underlying electronic structure.

\begin{figure}
     \includegraphics[width=0.45\textwidth]{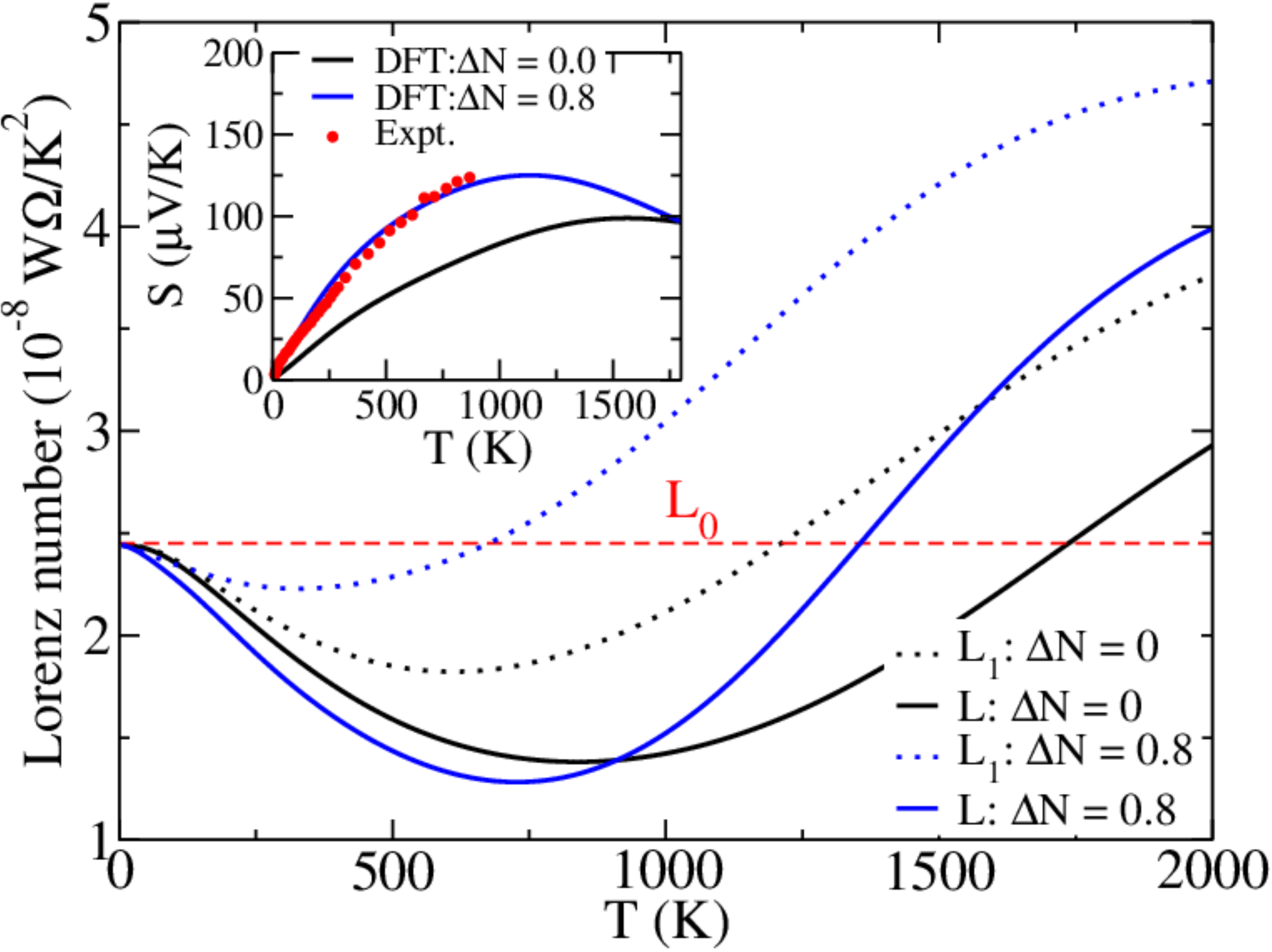} 

    \caption{
DFT results for Ba$_8$Au$_{6}$Ge$_{40}$:    
Coefficients $L$ (full lines) and $L_1$ (dotted lines) of the Lorenz tensor as a 
function of temperature for no doping 
(black)  and for a doping of $\Delta N = 0.8$ electrons (blue). 
The constant Lorenz number L$_0$ of the Wiedemann-Franz law is indicated by a dashed
horizontal line in red.  The inset shows the corresponding DFT-derived Seebeck
coefficients in comparison to the experimental values of Ref.
\cite{zeiringer_phase_2011}.
}
 \label{fig2}
  \end{figure}

Fig.~\ref{fig1}(a) shows the density of states (DOS) around Fermi
energy. A gap of about 0.3 eV occurs about 0.15 eV above the Fermi energy
$E_F$ for the undoped case. Electron dopings of up to $\Delta N =1.2$ places
$E_F$ closer to the gap where the DOS diminishes rather strongly. For such a
situation a large Seebeck coefficient is expected, as is indeed the case (see inset in Fig. \ref{fig2}).  
Assuming two valence electrons for each Ba atom, one valence electron
for each Au atom and for valence electrons for each Ge atom, 
the total number of valence electrons of
Ba$_8$Au$_{6}$Ge$_{40}$ amounts to 182.
Finally, by adding $\Delta N = 2$ electrons 
the corresponding $E_F$ falls into a 
gap because then the total number of valence electrons
is 184, which is precisely the total number of valence electrons
of Ge$_{46}$ in its clathrate structure, 
which at Fermi energy has a rather large gap of about 1.6 eV.
This is a remarkable feature since the
electronic structure of Ge in its
diamond ground state structure reveals no gap at all when 
a local approximation of the DFT exchange-correlation interactions
is utilized, as it is the case here. Such an underestimation of
the gap size  is a well-known shortcoming of standard DFT calculations. Assuming that the
gap is preserved when filler atoms such as Ba 
are placed in the
voids of the clathrate structure and elements such as Au 
are substituting Ge according
to the composition Ba$_8$Au$_{6-x}$Ge$_{40+x}$, then the relation
$x=\Delta N / 3$ can be established. The critical composition, i.e. the Fermi
energy falls into the gap, would then be $x_{crit.}=2/3$. 

Fig.~\ref{fig1}(b) depicts the coefficients $L$ and $L_1$ of the cubic Lorenz tensor
(see Eqs. \ref{L_tensor} and \ref{L1_tensor}) at 300 K as
functions of doping. 
The inset shows the carrier-density-dependency of the resistivity
$\rho$ which is the inverse of the conductivity $\sigma$ in Eq.~\ref{electrical_conductivity}.
Significant deviations between the Lorenz coefficients occur in the range of $0 \leq \Delta N \leq 4$,
which increase strongly when the doping is near to $\Delta N =2$.
At $\Delta N = 2$ the coefficients $L$ and $L_1$ are  undefined because in the gap
no electronic states are available for transport and therefore at
lower temperatures 
the electrical conductivity $\mathbf{\sigma}$ is zero (or the resistivity has
a pole, see inset in Fig.~\ref{fig1}(b) ), 
and as a consequence the Lorenz coefficients reveal singularities.
The comparison between $L_1$ and $L$ demonstrates the influence of the
Seebeck coefficient in terms of  -$S^2$, 
which increases dramatically as the Fermi level approaches the
gap upon doping.

%change2
For thermoelectric applications, both  $S$ as well as  $\mathbf{\sigma}$
should be large for obtaining  a large power factor $S^2\mathbf{\sigma}$.
For such a purpose, a suitable carrier concentration is required.
In the case of Ba$_8$Au$_{6-x}$Ge$_{40+x}$, a doping $\Delta N$ up to 1.2 
yields a carrier concentration of about $n=-2 \times 10^{20}$ e/cm$^3$, which
is within the desired range for thermoelectric properties. 
The negative sign of $n$ refers to hole carriers, i.e., Fermi energy is below
the gap, a positive sign just refers to the opposite case, i.e., Fermi energy
above the gap.

For dopings $\Delta N > 4, \Delta N < 0$ the coefficients $L$ and $L_1$ nearly coincide 
and their values become comparable to
the constant Lorenz number but  $L_0$ could still be off by 20\%.
The results suggest that the approximation by $L_1$ is not
useful for thermoelectric materials with large Seebeck coefficients.

Fig.~\ref{fig2} reveals the temperature dependency of  $L_1$ and $L$ for no doping
and $\Delta N=0.8$.  The inset compares DFT--derived and experimental Seebeck
coefficients revealing very good agreement between measurement and DFT
calculation for a doping of $\Delta N=0.8$. Such a doping refers to an
off-stoichiometry of $x=0.27$, which is within
the experimental error measuring the composition \cite{zeiringer_phase_2011}.
Both $L_1$ and $L$ approach the WF limit $L_0$ at low temperatures but deviate
significantly at elevated temperatures.  Overall, $L$ and $L_1$ exhibit a
strong temperature dependency and the deviation of the constant value $L_0$
becomes large in particular around 700 K, which is in the temperature range of
technological applications. 

%change1 
As demonstrated by the inset of Fig.\ref{fig2} the assumption of a
constant relaxation time $\tau$ works well for the evaluation of the Seebeck coefficient.
Nevertheless, we tested an energy dependent relaxation time  
$\tau(\varepsilon)=const \,\, \varepsilon^{-1/2}$  according to Ref.~\onlinecite{blatt_1968} 
recalculating $L$ as well as $S$.
This ansatz for $\tau(\varepsilon)$
is supposed to model the scattering of electrons by acoustic phonons. The
energy $\varepsilon = |E_F - \varepsilon_{\vec{k},\nu}|$
was defined as the absolute value of the difference of
Fermi energy and the respective band energy for band $\nu$ and vector
$\vec{k}$.

  \begin{figure}
     \includegraphics[width=0.45\textwidth]{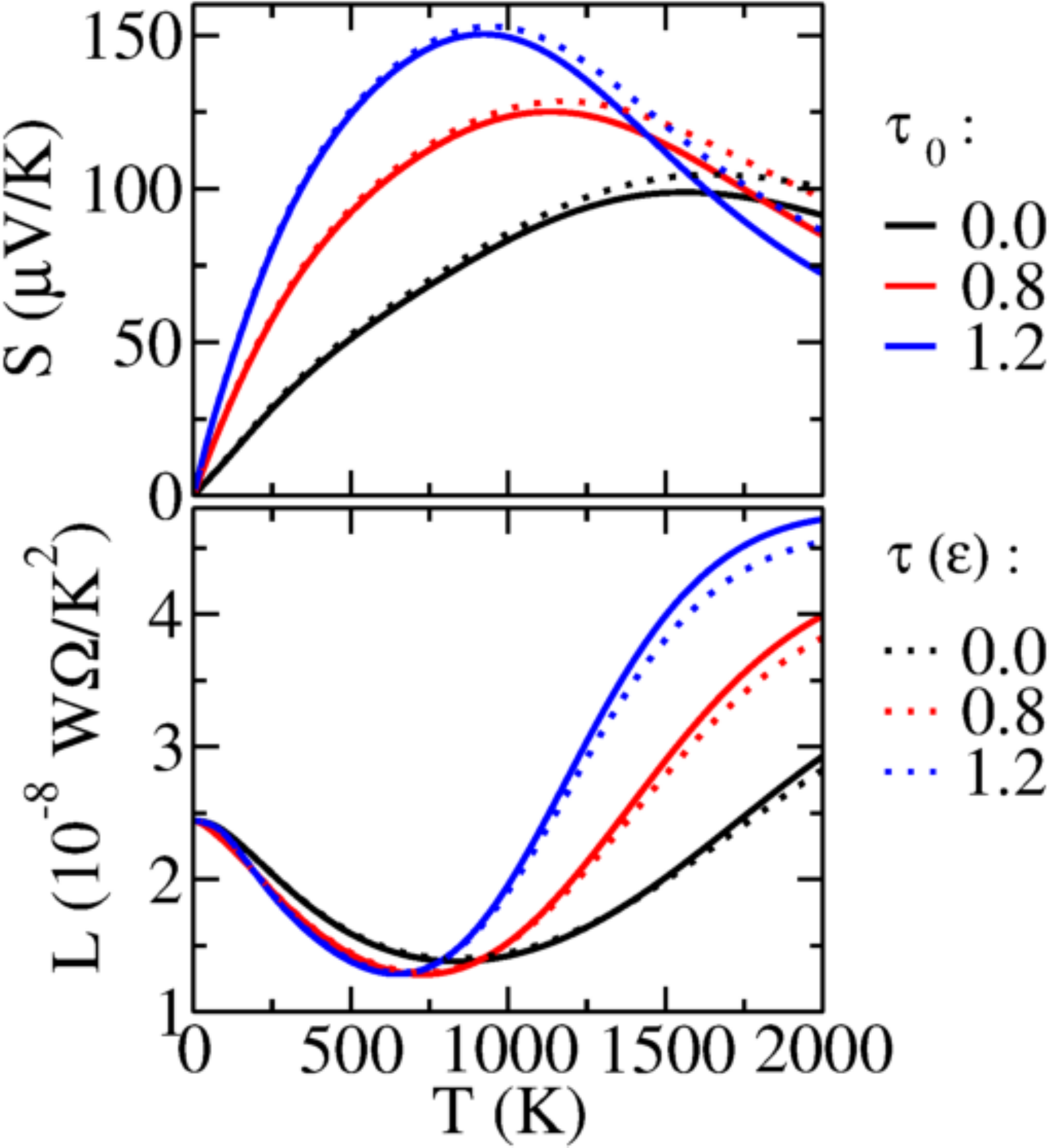}
    \caption{DFT results for Ba$_8$Au$_6$Ge$_{40}$:
Seebeck coefficient $S(T)$ and the component
$L(T)$ of the full Lorenz tensor for no doping and dopings of $\Delta N=0.8,
1.2$.
Results with constant relaxation time $\tau_0$ (full lines)  and with the
energy-dependent ansatz 
$\tau(\varepsilon)= const \,\, \varepsilon^{-1/2}$ (dotted lines). 
}
 \label{fig3}
  \end{figure}

By considering this specific  $\tau(\varepsilon)$, based on the energy dispersions of 
the two-band Kane approximation 
Huang \textit{et al.} demonstrated for Bi$_2$Te$_3$\cite{huang_ab_2008},
that a constant relaxation time yields a slightly larger Lorenz number than the energy
dependent ansatz. We performed similar calculations for Ba$_8$Au$_6$Ge$_{40}$
for the dopings $\Delta N =0, 0.8, 1.2$ by utilizing the DFT-derived electronic structure.
In this context, it should be noted that for deriving the relation
$\tau(\epsilon) \propto \epsilon^{-1/2}$ bands with parabolic
dispersions near Fermi energy are assumed, which is not really the case for
the material under study. Nevertheless, the results for $S(T)$ and $L(T)$
in Fig. \ref{fig3} show that the influence of the assumed energy dependency of
$\tau(\varepsilon)$  is rather small, although it becomes more pronounced at elevated
temperatures and larger dopings with Fermi energy approaching the gap.

  \begin{figure}
     \includegraphics[width=0.45\textwidth]{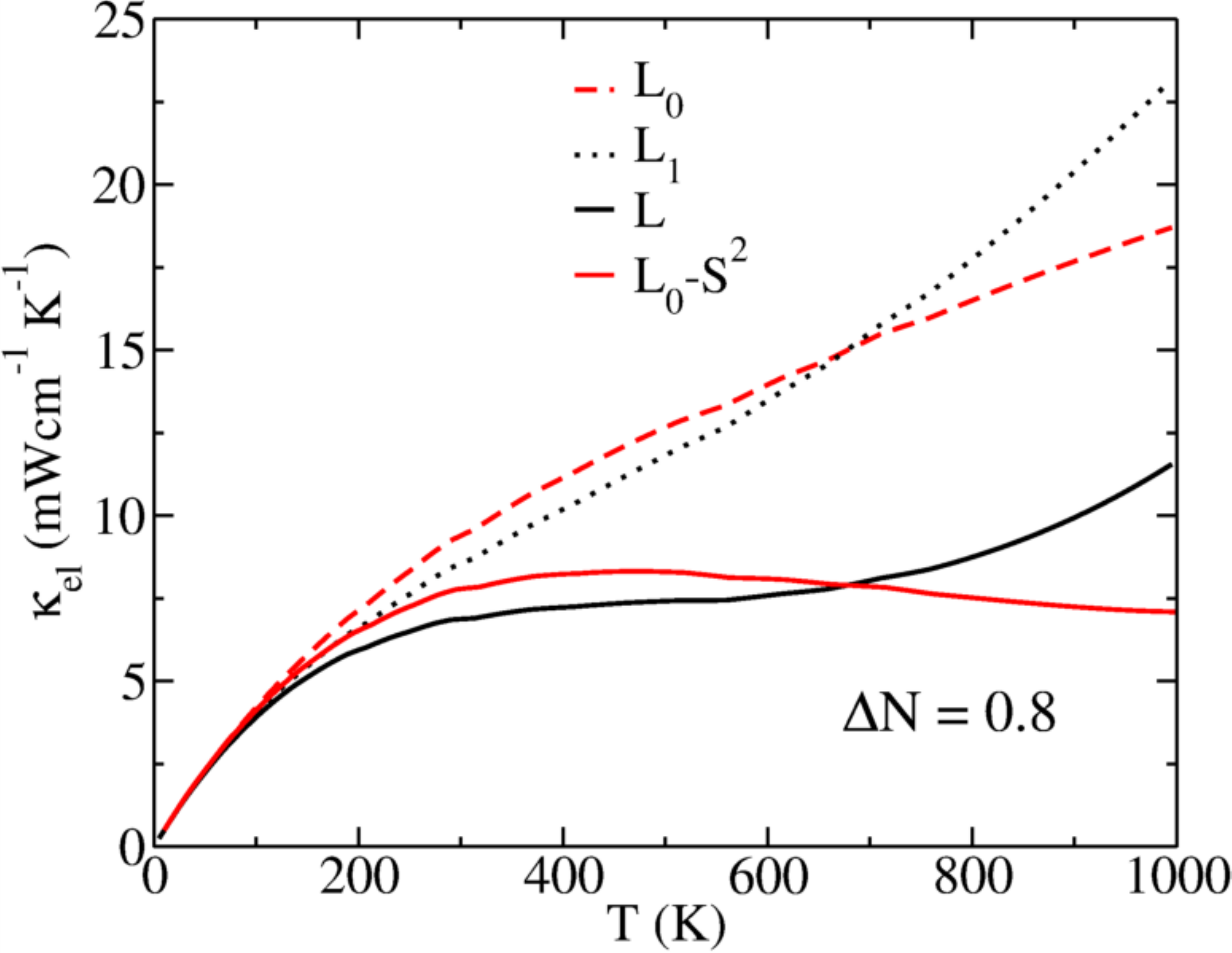}
    \caption{DFT derived electronic thermal conductivities $\kappa_{el}$
for Ba$_8$Au$_{6-x}$Ge$_{40+x}$. Results for a doping of $\Delta N=0.8$ using the approximations
$L_0$,$L_1$,$L_0-S^2$ and $L$ of the full calculation. $S$ is the Seebeck coefficient.
}
\label{fig4}
  \end{figure}

For finally deriving the electronic thermal conductivity $\kappa_{el}$ from
Eq.~\ref{kappa_wf} the full Lorenz coefficient $L$ together with its
approximations $L_1$ and $L_0$ are used.  The results in Fig.~\ref{fig4}
reflect the behavior of the Lorenz coefficients in Fig.~\ref{fig2} showing
again  very significant deviations up to 40\% between the full calculation and
the result involving the approximations $L_1$ and $L_0$. Clearly, just using
the simple WF law (i.e. $L_0$) may lead to rather unreliable values for the
electronic thermal conductivity, and consequently for the lattice thermal
conductivity when it is derived from the measured total thermal conductivity
by $\kappa_{ph}=\kappa^{meas.} - \kappa_{el}$.  The rather similar behavior of
thermal conductivities $\kappa_{el}$ in Fig.~\ref{fig4} as calculated 
with  $L$ and the approximation
$L_{app.}=L_0-S^2$ would suggest to use $\kappa_{el} \approx T L_{app} \sigma$ for
a reasonable estimation of $\kappa_{el}$. This would have the big
advantage, that $\kappa_{ph}$ could be estimated from measurable
quantities, namely the Seebeck coefficient $S(T)$ and the electrical
conductivity $\mathbf{\sigma}$. This observation, however, is only useful as
long as $S^2$ is not too large and temperatures are sufficiently.

  \begin{figure}
     \includegraphics[width=0.45\textwidth]{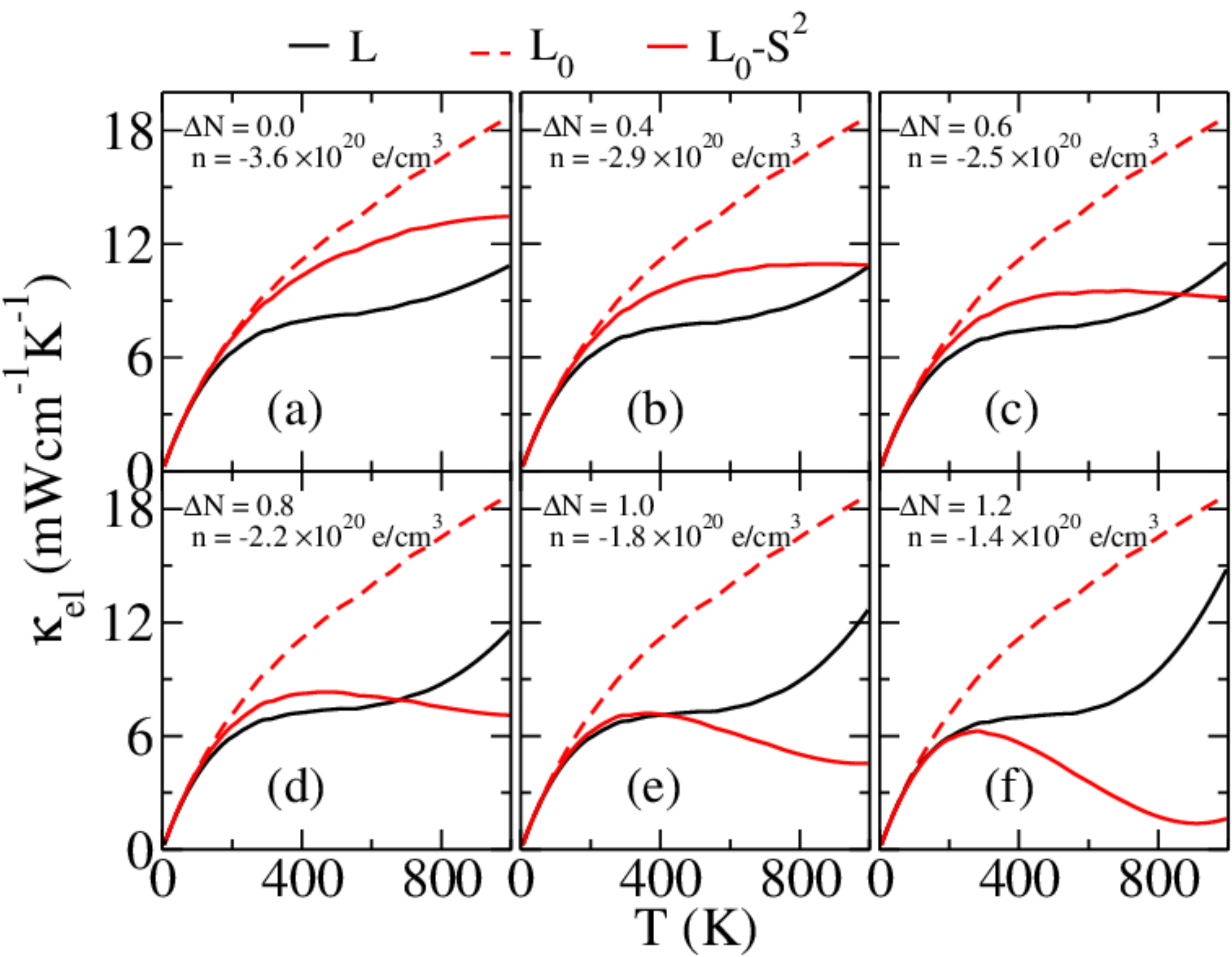}
    \caption{DFT derived electronic thermal conductivities $\kappa_{el}$
for Ba$_8$Au$_{6-x}$Ge$_{40+x}$. 
Results for a series of dopings $\Delta N$ using $L_0$,$L_0-S^2$, and
$L$ as Lorenz coefficients.  The corresponding carrier concentrations are also
indicated.
}
 \label{fig5}
  \end{figure}

As Fig.~\ref{fig5} however reveals, for dopings $\Delta N \geq 0.8$ the use of the
approximation $L_0-S^2$ grossly underestimates the results
obtained with $L$, the Lorenz coefficient calculated without any
approximation (apart from the assumption of a constant
relaxation time). The errors become larger the nearer $E_F$ approaches the gap.
This is attribute to the fact that $L_1$ deviates $L_0$ significantly 
when the system becomes semiconducting/insulating as shown in the inset of Fig.~\ref{fig1}(b). 
Likewise, only using the constant $L_0$ of the Wiedemann-Franz results in
significant overestimations. Therefore the conclusion has to be made,
that it is mandatory to derive the electronic thermal
conductivity from DFT calculations in combination with
Boltzmann's transport theory in terms of the full Lorenz-tensor
$\mathbf{L}$. Otherwise the estimated $\kappa_{el}$ and,
subsequently, $\kappa_{ph}$ might be rather wrong, which is
undesirable for optimizing thermoelectric properties of a
material.

\begin{acknowledgments}   
The authors gratefully acknowledge the support by the Austrian Science
Foundation FWF under project nr. P24380-N16.
The DFT calculations were done on the Vienna Scientific Cluster (VSC).
\end {acknowledgments}

\appendix*
\section{}
The tensor  needed in Boltzmann's transport theory is

\begin{eqnarray} 
\mathbf{K_n} &=& \frac{1}{4\pi^3}\sum_{i,\mathbf{k}}\tau_i(\mathbf{k})\mathbf{v}_i(\mathbf{k})
\otimes
\mathbf{v}_i(\mathbf{k})(\varepsilon_i(\mathbf{k})-\mu)^n  \\ \nonumber
 &&  \left(-\frac{\partial f(\mu,T,\varepsilon_i)}{\partial \varepsilon_i}\right)
  \label{kn}
\end{eqnarray}

in which $\tau_i(\mathbf{k})$ is the relaxation time of the electronic states
with band index $i$, energy eigenvalue $\varepsilon_i$ and  band velocities
$\mathbf{v}_i$  for  wave vector $\mathbf{k}$, whereas
$f(\mu,T$ denotes the Fermi-Dirac distribution function for the chemical potential $\mu$ at
temperature $T$. 

The electrical conductivity tensor $\mathbf{\sigma}$ is 
given by
 
\begin{equation}
\mathbf{\sigma} = e^2\mathbf{K_0}
\label{electrical_conductivity}
\end{equation}

and the Seebeck tensor $\mathbf{S}$ is defined by 

\begin{equation}
\mathbf{S} = \frac{1}{eT}\mathbf{K_1 K_0}^{-1} .
\label{seebeck}
\end{equation}

\bibliographystyle{apsrev4-1}
\bibliography{references}

\end{document}